\documentclass[a4paper]{aa}
\usepackage{graphicx}
\usepackage{natbib}

\newcommand{\NBpc}{\mbox{NB$_-$3.21}}
\newcommand{\NBp}{\mbox{NB$_-$3.28}}

\def\aap{A\&A}

\def\apj{ApJ}
\def\apjl{ApJL}
\def\apss{ApSS}
\def\nat{Nature}
\def\mnras{MNRAS}
\begin{document}

   \title{New insights on the complex planetary nebula Hen 2-113\thanks{Based on observations
       made with the Very Large Telescope Interferometer at
       Paranal Observatory under programs 073.D-0130, 074.D-0139}}

   \author{E.~ Lagadec \inst{1}, O.~Chesneau\inst{2}, M.~Matsuura\inst{3},
   O.~De Marco\inst{4}, J.A. de Freitas Pacheco\inst{1}, A.A. Zijlstra\inst{3},
   A. Acker\inst{5}, G.C. Clayton\inst{6}}

   \offprints{E.Lagadec,email:lagadec@obs-nice.fr}
\institute{Observatoire de la C\^{o}te d'Azur-CNRS-UMR 6202, Dept. Cassiop\'{e}e,
BP 4229, F-06304 Nice, France
\and
Observatoire de la C\^{o}te d'Azur-CNRS-UMR 6203, Dept. Gemini,
Avenue Copernic, F-06130 Grasse, France
\and
Department of Physics and Astronomy, University of Manchester, Sackville
Street, P.O. Box 88, Manchester M60 1QD, UK
\and
American Museum of Natural History, Dept. of Astrophysics,
Central Park West at 79th Street, New York
NY 10024, USA
\and
Observatoire de Strasbourg, 11 rue de l'Universit\'{e}, 67000 Strasbourg, France
\and
Department of Physics and Astronomy, Louisiana State University, Baton Rouge, LA
70803, USA
}

   \date{Received; accepted }
\titlerunning{Hen~2-113}
\authorrunning{Lagadec, Chesneau, Matsuura et al.}
   \begin{abstract}{We report infrared observations of the planetary
nebula Hen 2-113 obtained with VLT/NACO, VLTI/MIDI, VLT/ISAAC and TIMMI at
the ESO 3.6m. Hen 2-113 exhibits a clear ring-like structure
superimposed to a more diffuse environment visible in the L' (3.8$\mu$m), M' (4.78$\mu$m) and
8.7$\mu$m bands. No clear core at 8.7$\mu$m and no fringes through
the N band could be detected for this object with MIDI. 
A qualitative interpretation of the object structure 
is proposed using a diabolo-like
geometrical model.
The PAH content of the nebula was also studied with ISAAC and TIMMI
observations. This indicates that the PAHs are mostly concentrated towards the
lobes of the diabolo and the bipolar lobes of the nebula.
In L' band, a void $0.3\arcsec$ in diameter was discovered with NACO around the central
source. The L' and M' fluxes from the central source were derived from NACO
data indicating an important infrared excess with respect to the expected
stellar emission based on stellar models and short wavelength data. The
observed flux from this  source in the L' and M' is about 300 and 800
times respectively than those expected from a model including only the central
star. Moreover, the central object appears resolved in L' band with measured
FWHM about 155 mas. This infrared
excess can be explained by emission from a cocoon of hot dust (T$\sim$1000K)
with a total mass $\sim10^{-9}$M$_{\odot}$. 

   \keywords{Techniques: interferometric; Techniques: high angular
                resolution; Stars: AGB and post-AGB ;individual: Hen~2-113;
                Stars: circumstellar matter; Stars: mass-loss; Stars:imaging; Infrared: stars;}
   }
\end{abstract}
   \maketitle
%

\section{Introduction}

One of the most debated questions regarding the post-AGB evolution of low and
intermediate mass stars is the departure from spherical symmetry observed in
circumstellar envelopes of Pre-Planetary Nebulae (PPNe) and Planetary nebulae 
(PNe). Indeed, whereas AGB stars have envelopes roughly spherical, many PNe
exhibit axisymmetric structures or even more complex morphologies.

Many theoretical models have been proposed to explain shapes of PPNe and PNe (see 
a recent review by Balick \& Frank 2002). Most PNe and PPNe shaping theories rely
on the velocity field of circumstellar material distributed either in an expanding toroidal 
structure or in an accretion disk. The formation of such structures requires the presence 
of a binary companion, rotation and/or magnetic fields. In past years, various
observations have been conducted aiming to detect features associated with mechanisms
responsible for the shaping of PNe. De Marco et al. (2004), as well as Sorensen
\& Pollacco (2004), showed that a significant binary population may be present among
the central stars of PNe. Detection of magnetic fields around AGB stars and
central stars PNe has also been mentioned recently by  different teams (Miranda et al. 2001, Etoka \&
Diamond 2004, Bains et al. 2004, Jordan et al. 2005), but the field strengths are not
enough to dominate the flow dynamics. Since the most extreme bipolar morphologies appear to be
related to circumstellar disks (Balick 1987), the detection and study of disks close to
central stars is a very active field of research. Presently, this
field benefits particularly from high spatial resolution techniques like
adaptive optics on 8m class telescopes (see, for instance, Biller et
al. 2005) and recently, from the advent of interferometry in the
mid-infrared (MIR) wavelength range (Leinert et al. 2004). These new facilities give a complementary view of 
circumstellar disks discovered by the HST (Sahai \& Nymann 2000, De Marco et al. 2002, Meakin et al. 2003). 
A good example of such an approach is the study of the disk in NGC 6302 by Matsuura et al. (2005), using 
HST and VLT observations. 


In this sense, the compact and young PNe Hen~2-113 (Hen~3-1044, PK 321+03.1, IRAS 14562-5406) 
is a good example of an object with a complex morphology which, in spite of its
kinematical properties, is typical of a type IIb planetary
nebula (Costa \& de Freitas Pacheco 1996), this type of PNe being mostly
circular in shape (Phillips 2005). The central star (CS) has
a Wolf-Rayet type spectra and was classified as a [WC10] by 
Crowther, De Marco \& Barlow (1998). CS exhibiting WR spectra
represent about 10-15\% of  all CS (G\'orny \& Tylenda, 2000).

The ISO spectra of Hen~2-113 show simultaneously
the presence of C-rich and O-rich dust grains (Waters et al. 1998). 
The dual dust chemistry phenomenon in PNe appears to be correlated
with the presence of cool [WC] type CS (Zijlstra et al., 1991; De Marco \& Soker 2002).

De Freitas Pacheco et al. (1993) carried out a spectral analysis of the wind of
Hen~2-113, concluding that the carbon-to-helium ratio is C/He = 0.5, value
similar to the value of 0.55 found in an independent way by De Marco \&
Crowther (1998). They also derived a 
mass-loss rate of $3.8\times 10^{-6} $M$_{\odot}$yr$^{-1}$ for an estimated
terminal velocity of 1000 km s$^{-1}$. If the terminal velocity is revised downward
($V_{\infty} \sim 250$ km $ $s$^{-1}$), the resulting mass-loss rate would be decreased
by a factor of 8, since $\dot M \propto V_{\infty}^{3/2}$. This revised value is consistent with the
estimate by Leuenhagen, Hamann \& Jeffery (1996) of $3.4\times 10^{-7}
$M$_{\odot}$yr$^{-1}$ and smaller than the estimation of $3.9\times 10^{-6}
$M$_{\odot}$yr$^{-1}$ by De Marco \& Crowther (1998).   

From high resolution Coud\'e spectra, de Freitas Pacheco et al. (1993) decomposed the
main interstellar components of the NaI doublet and, using a rotation model for the Galaxy,
they have derived a distance of 3.1 kpc for Hen~2-113. A similar value (3.5 kpc) was obtained by
Leuenhagen, Hamann \& Jeffery (1996), but a lower distance (1.2 kpc) was
derived by De Marco \& Crowther (1998). The effective temperature of the CS was estimated to be
29000 K by De Marco \& Crowther (1998), who modeled the star atmosphere. From the
the H-Zanstra and the Stoy methods, de Freitas Pacheco et al. (1993) derived a lower effective
temperature, namely, $T_{eff}$ = 22300 K, consistent with the absence of HeII lines in the
wind and the low excitation nebular spectrum (de Freitas Pacheco, Maciel \& Costa 1992; De Marco,
Barlow \& Storey 1997).

Despite the careful study of the fundamental parameters of the CSs
and nebulae of Hen~2-113 and CPD-568032  by De Marco et al. (1997, 1998) and De Marco \& Crowther (1998),
the morphology of the nebula of Hen~2-113 was poorly known until
the HST observations by Sahai et al. (2000) (hereafter SNW00). These authors showed that
Hen~2-113 exhibits a complex geometry, roughly bipolar with two bright,
knotty, compact ringlike structures around the central star. This compact structure is embedded in a larger and fainter spherically symmetric structure
and SNW00 were puzzled
to notice that the central star of Hen~2-113 was conspicuously offset from the geometrical centers of
the rings and from  the circular contours attributed to the formal AGB wind.

In the present work we report new observations on Hen~2-113 performed with different infrared
imaging instruments (ISAAC and TIMMI) and high spatial resolution techniques, namely adaptive
optics with NACO/VLT. The
infrared domain offers a unique opportunity to observe with great detail
the bulk of the dust emission in the very central region of the object. In particular, we attempted to detect and study small scale 
structures in the MIR with the long baseline interferometer MIDI/VLTI. The object was over-resolved with 46m baselines and no interferometric data could be recorded, but the
acquisition images are used in our study.
The remainder of this paper is organized as follows: in Sect.2 the infrared observations are presented, in
Sect.3 we perform a comparative analysis between the infrared and optical HST images, in
Sect.4 a possible geometry for Hen~2-113 is discussed and, finally, in Sect.5
the conclusions are given.

\section{Observations}


\begin{figure*}
 \centering

\includegraphics[height=5.6cm]{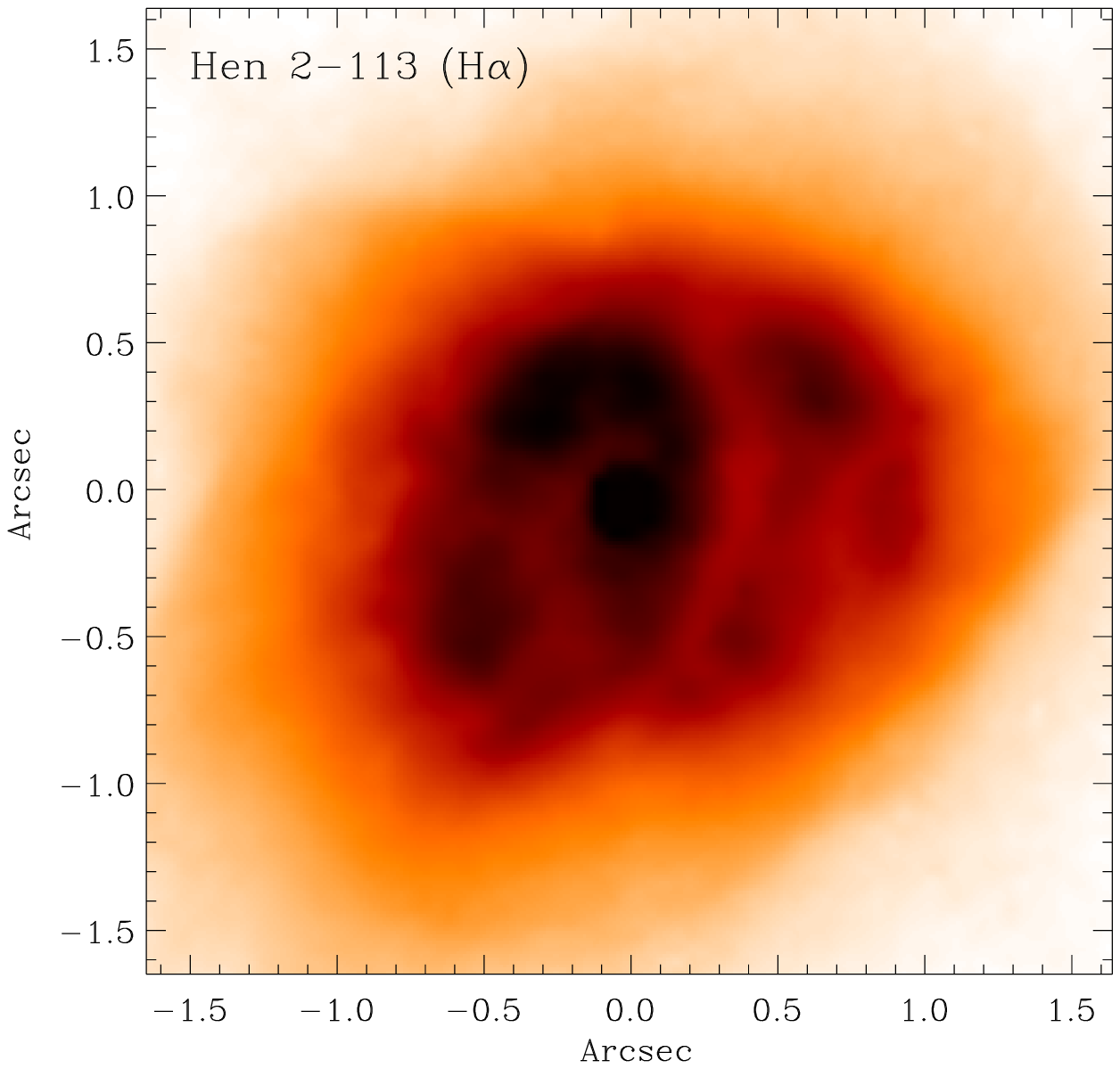}
\hspace{-2.5cm}
\includegraphics[height=5.6cm]{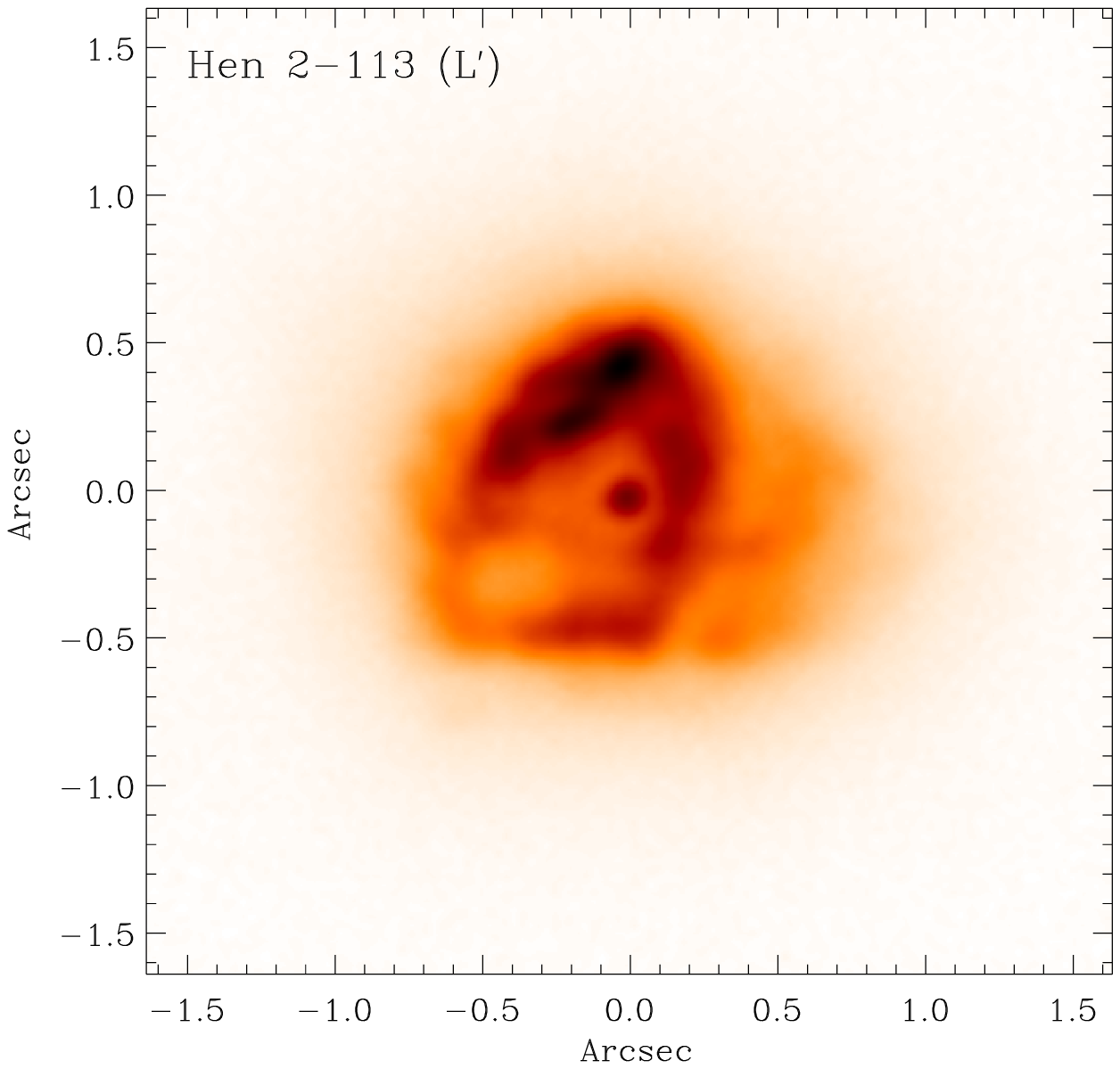}
\hfill
\includegraphics[height=5.6cm]{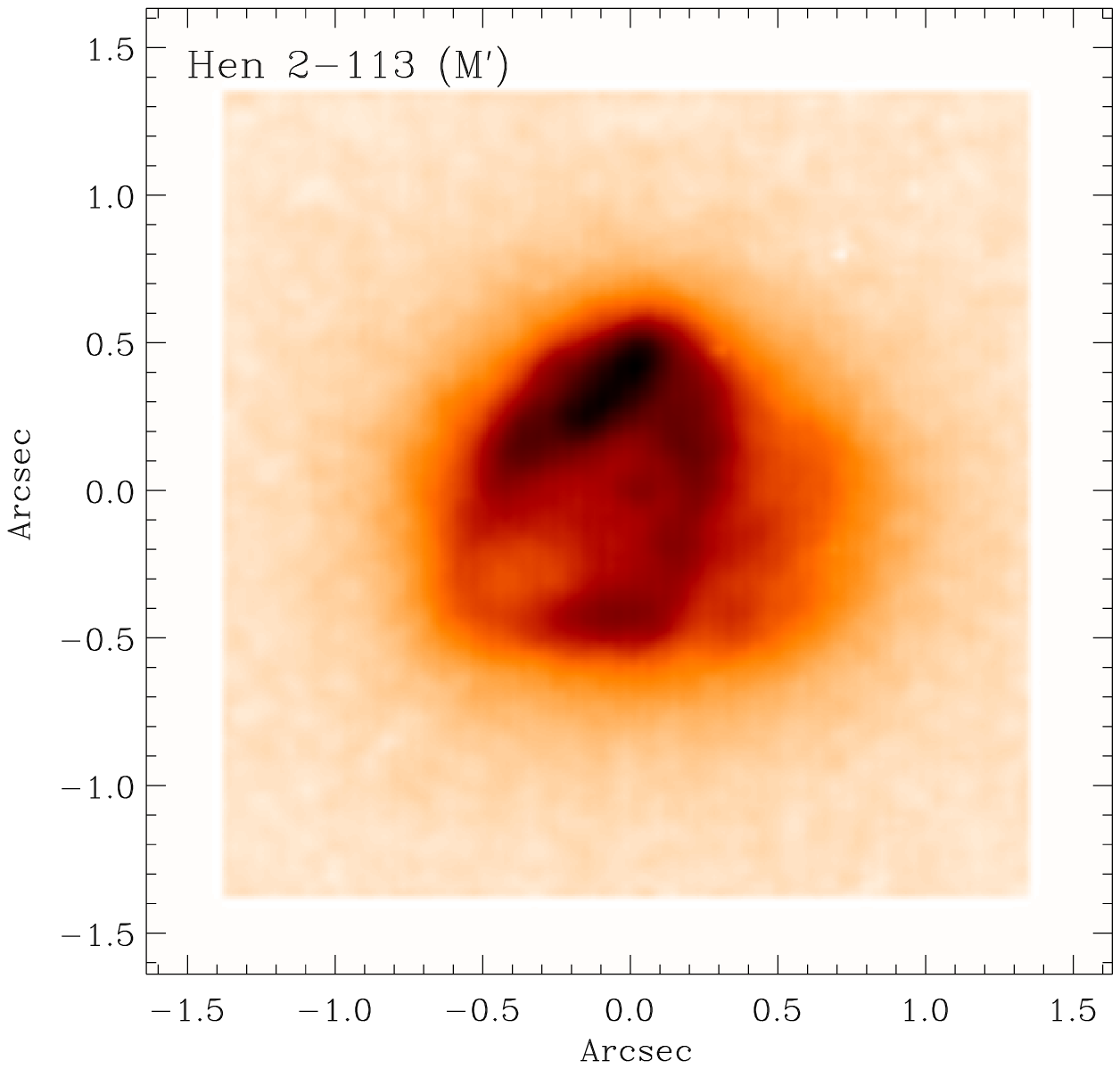}
\hspace{-2.5cm}
\includegraphics[height=5.6cm]{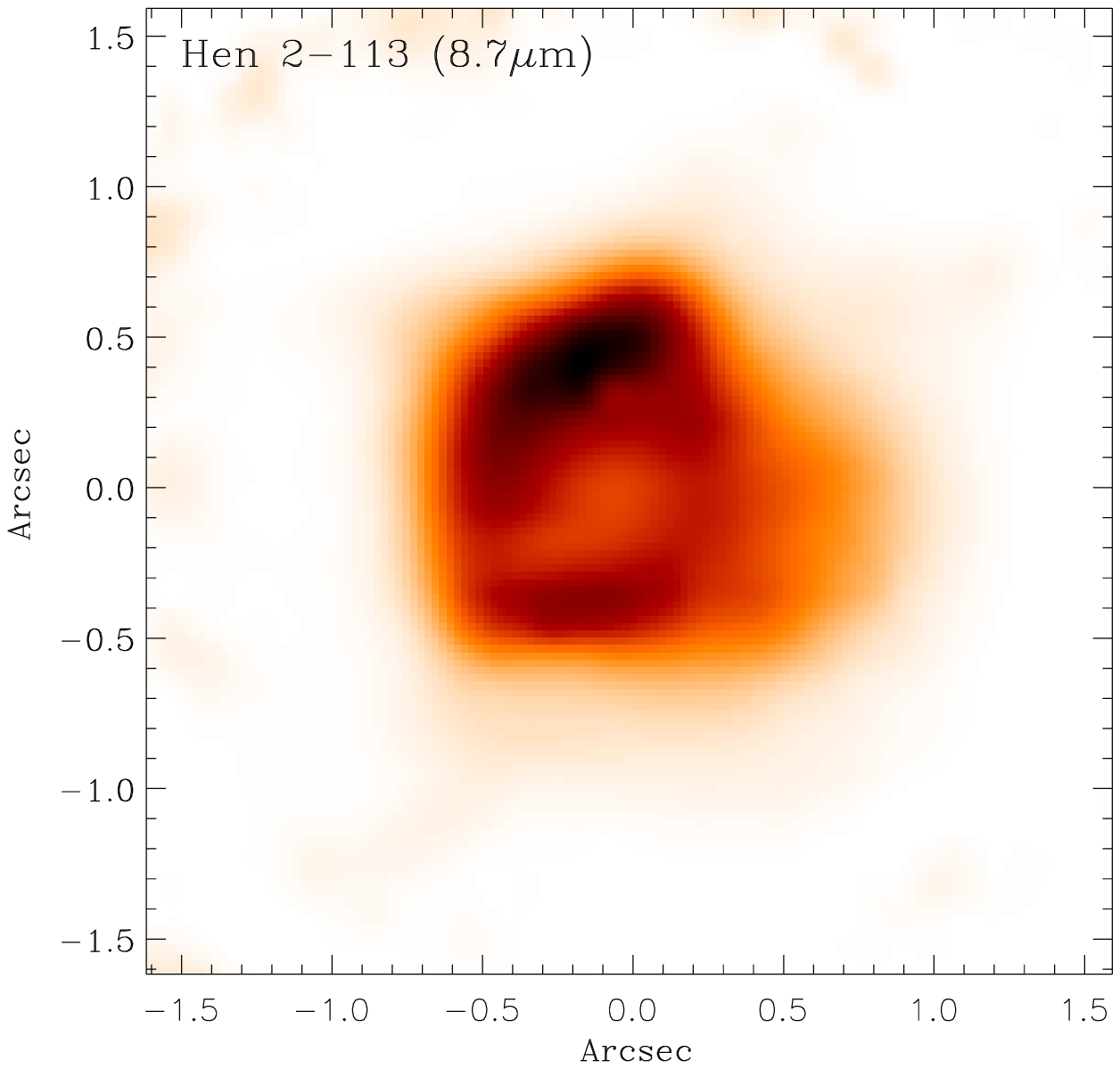}


 \caption[]{Comparison of the HST image in H$\alpha$ (upper
 left), the NACO L'(upper right) and M' (lower left), and the deconvolved MIDI
 acquisition image at 8.7$\mu$m. North is up and east to the left.
\label{fig:compimage}}
\end{figure*}

\subsection{NACO high resolution imaging}

We have observed Hen~2-113 with the adaptive optics camera NACO
attached to the fourth
8.2 m Unit Telescope (UT) of the Very Large Telescope (VLT), European Southern Observatory (ESO) 
Paranal, Chile. NAOS was
operated in the visual wavefront sensor configuration with the
SBRC Aladdin 1024$\times$1024 detector. We observed the target with L$'$ ($3.8 \mu$m)
and M$'$ ($4.78 \mu$m)
broad-band filters. Using camera mode L27, the field of view was
28$\arcsec$$\times$28$\arcsec$ and the pixel scale was 27.1~mas per
pixel. The auto-jitter mode was used, that is, at each exposure, the telescope
moves according to a random pattern in a 10$\arcsec$ box.
A cross-correlation technique was used to recenter the images at about 0.25
pixel accuracy.

\begin{figure}
  \begin{center}
\includegraphics[width=10cm]{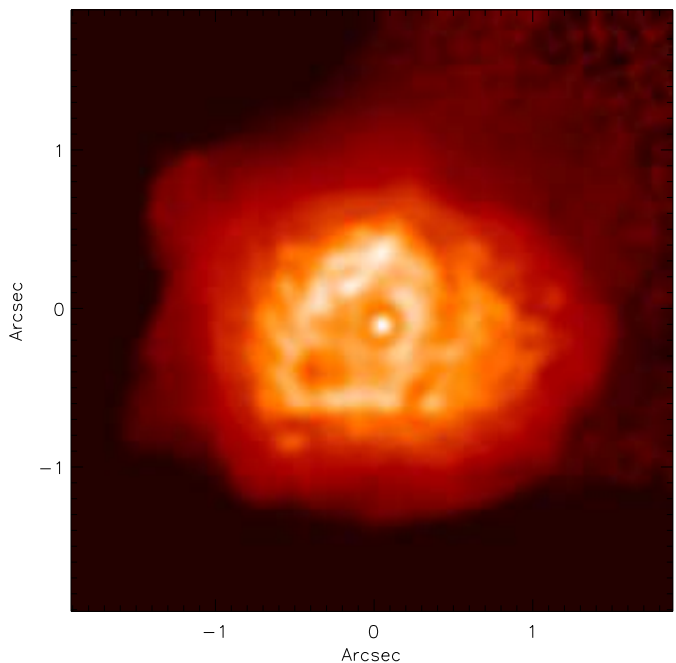}
  \end{center}
 \caption[]{NACO/VLT L' deconvolved image of Hen 2-113, log scaled in order to enhance the
   contrast. Dynamic range is $\sim10^4$ and resolution $\sim60$mas. In the North/west, low level
   artefacts of deconvolution are visible. North is up  and east to the left. 
\label{fig:L_image}}
\end{figure}

Individual dithered exposures were co-added, resulting in a total
exposure time $t_{\rm exp}$ shown in Table~\ref{tab:NACO}. The
data reduction was performed using an IDL
routine developed by us that processes the L' and M' individual frames as follows.
First, for L' images, bad pixels are removed. Then, the sky is
computed as the mean of the dithered exposures, and subtracted
frame by frame. Finally, all the sky-subtracted frames are shifted
and added together. The reduced broad-band images are shown in
Fig.~\ref{fig:compimage}. For the M' chopped images, we subtracted the
sky. The spatial resolution of our L' image has been improved by a deconvolution
with the Lucy-Richardson algorithm to reach $\sim$ 60 mas (Fig. \ref{fig:L_image}). 

\begin{table}[h]
\caption{\label{tab:NACO}Journal of observations with NACO/UT4.}
\vspace{0.3cm}
\begin{center}
\begin{tabular}{lllccc}
\hline
\hline
Star & Filter & Camera & Time &$t_{\rm exp}$ & Seeing\\
\hline
\multicolumn{6}{c}{04/05-05-2004, airmass$~$1.15-1.3}\\
\hline
Hen 2-113$^i$ & M$'$ &L27&02:34:48&37s&0.74\\
HD 130572$^i$& M$'$ &L27&03:19:34&18s&1.1\\
\hline
\multicolumn{6}{c}{13/14-05-2004, airmass$~$1.15-1.20}\\
Hen 2-113$^i$& L$'$ &L27&02:08:30&100s&0.42\\
\hline
\multicolumn{6}{c}{29/30-06-2004, airmass$~$1.15-1.20}\\
Hen 2-113$^i$& L$'$ &L27&05:17:54&40s&0.90\\
\hline
\multicolumn{6}{c}{04/05-07-2004, airmass$~$1.15-1.20}\\
\hline
Hen 2-113$^i$& L$'$ &L27&23:46:22&40s&0.93\\
Hen 2-113& L$'$ &L27&23:59:41&40s&0.76\\
HD 130572$^i$& L$'$ &L27&00:21:31&200s&0.66 \\
\hline
\end{tabular}
\end{center}
$^i$ with neutral density filter
\end{table} 

The star HD 130572 (A0V) was used to derive the point spread function (PSF) and as photometric 
standard. Its J, H and K magnitudes, from the 2MASS catalogue, are ,respectively, 6.39$\pm$0.02, 6.41$\pm$0.02, 
6.37$\pm$0.02 and its FWHM is 114$\pm$4~mas and 147$\pm$6~mas in the L' and M'
bands, respectively. The magnitude in the L' and M' bands was
assumed to be 6.39. 

\begin{figure*}
  \begin{center}
\includegraphics[width=8cm]{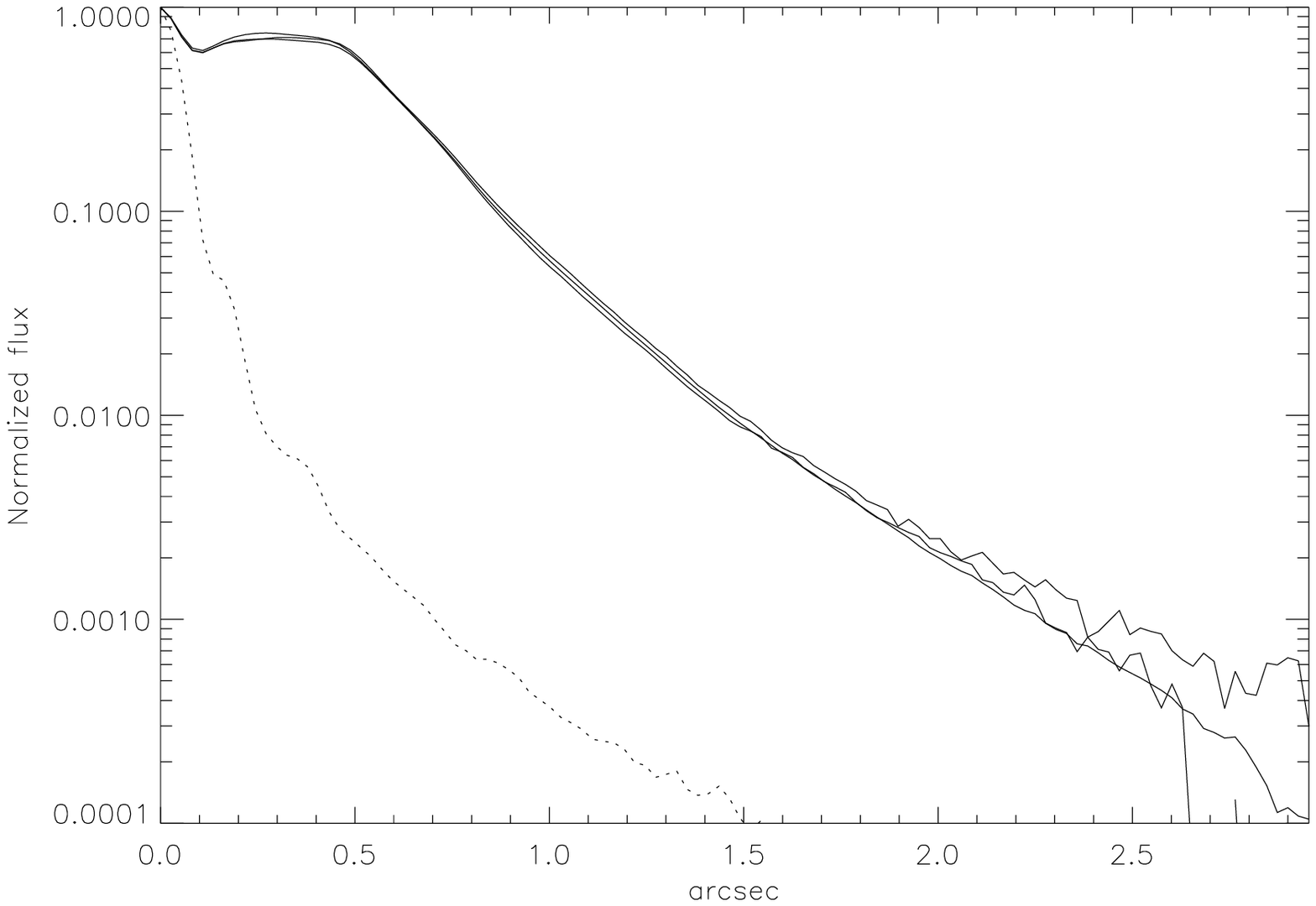}
\includegraphics[width=8cm]{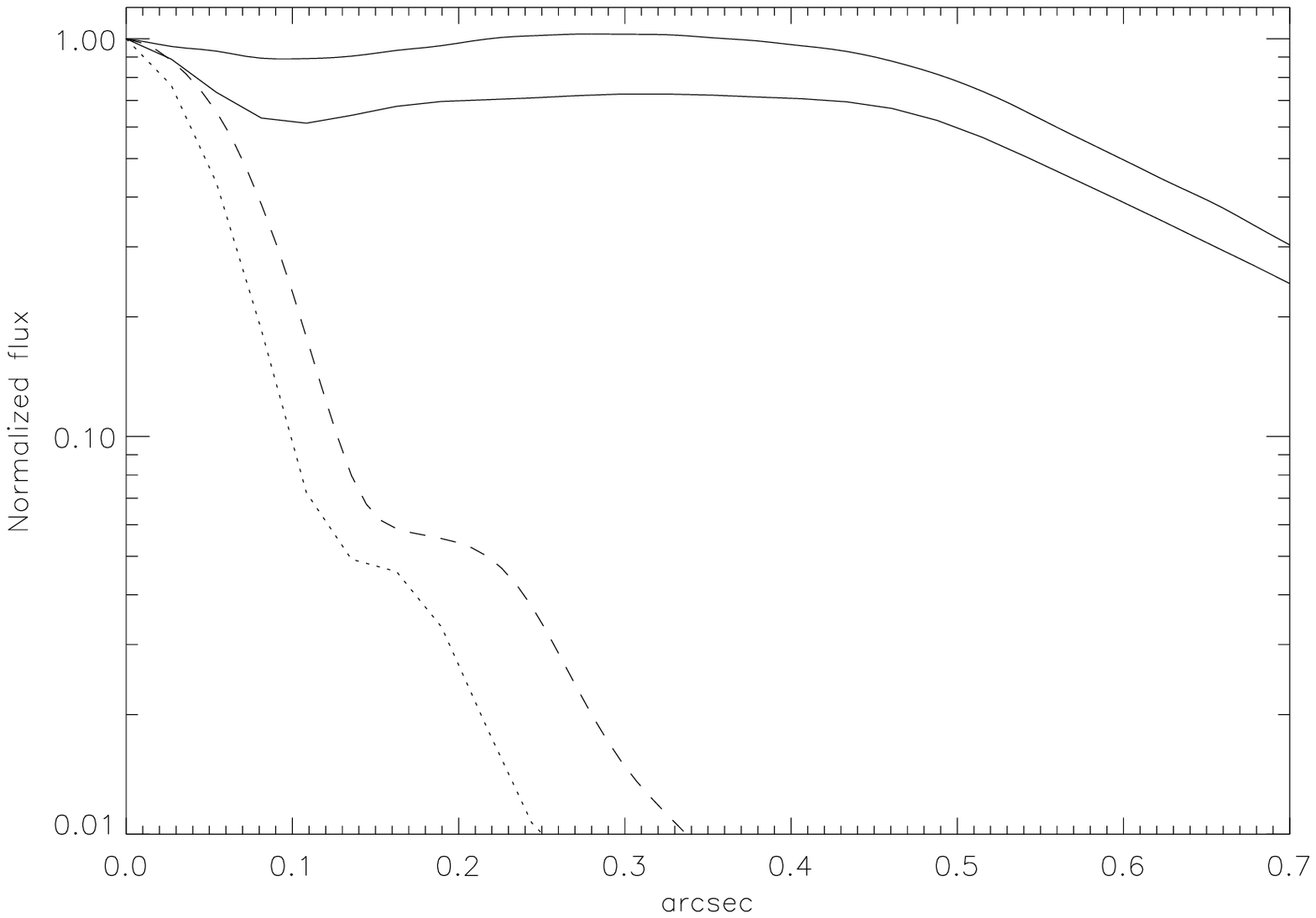}
  \end{center}
 \caption[]{Left: Radial flux normalized to the peak (filter L') for the three
   different observations of Hen 2-113 (solid
   line), and the PSF HD 130572 (dotted line). Right: closer view of the L' (dotted) and M' (dashed) PSF curves
   compared to Hen 2-113 L' (lower) and M' (upper).
\label{fig:radcavity}}
\end{figure*}

We estimated the magnitude of the full nebula to be $m_{L'}$=4.1$\pm$0.2 and
  $m_{M'}$=3.2$\pm$0.3\footnote{The quality of the M' images is quite limited in
    terms of noise and dynamic range leading to the large error bar for the
    integrated flux of the full nebula.} in the L' and M' bands respectively. 
Then, we performed photometry of the central source using a method described in Chesneau et al. (2005), based on PSF subtraction techniques.
The L' and M' radial means of the PSF and the source were computed, and the L' curves are shown in Fig.~\ref{fig:radcavity}. The radial mean of the PSF was then scaled to the central object to evaluate its photometry, independently
of the neighboring nebula,
based on the assumption that this object is indeed unresolved. 
The L' and M' magnitudes of the central source of 
Hen~2-113 were estimated as $m'_{L'}$=8.4$\pm$0.1 mag and $m'_{M'}$= 8.3$\pm$0.2 mag, respectively.
Interestingly, it appeared that the central source could not be considered as unresolved and could be fitted with 
an Airy curve having a FWHM of 155$\pm$16~mas, i.e. close to the M' band
  spatial resolution. We stress that this result is based on the comparison of the radial curves
from the four data sets recorded in L' band. The curves agree within
about 4\% and 2\% at 0.27 and 0.54 arcsecond respectively and
span a seeing range from 0.42 to 0.93 arcsec. In the M' band, the central object is also extended and embedded in a diffuse environment (Fig.~\ref{fig:radcavity}).

\subsection{MIDI acquisition image}

The VLTI/MIDI interferometer  (Leinert et al. 2003 and
2004) operates like a classical Michelson
interferometer to combine the MIR light (N band, 7.5 -
13.5~$\mu$m) from two VLT UTs. Observations of Hen 2-113 and CPD -568032 have
been performed in 2004, April 08. The observations of CPD -568032 will be
presented in a forthcoming paper (Chesneau et al 2005 (in preparation)).
The log of the observations is presented in
Table~\ref{tab:journal}. For these observations, the UT2 and the UT3 telescopes were
used with a separation of 46.6~m and with the baseline oriented 40$^\circ$ (E
of N). This baseline provides a resolving power of the order of 40 mas and was adopted
in order to detect compact dusty structures around the central star.

The observing sequence, typical of interferometric measurements
is described in Leinert et al. (2003, 2004) and Przygodda et al. (2003). It must be stressed that the MIDI instrument
detected no fringes as the target appeared resolved for a single dish 8~m
telescope. Nevertheless, the acquisition images from this instrument were
scientifically interesting and their reduction is described in the next
Section.  

The chopping mode (f = 2 Hz, angle = -90$^\circ$) was used to visualize and to accurately point at
the star, which is usually not perfectly centered in a first acquisition
image, and is centered in a second step. Such a pointing accuracy is needed to
get good quality interference pattern. The number of frames
recorded for each image was generally 2000, and the exposure time
per frame is by default 4~ms to avoid fast background saturation.
If the pointing is not satisfactory, the procedure
is started again. The default filter is centered at 8.7$\mu$m
(1.6$\mu$m wide). This filter has been used for the three
acquisitions of the science targets and the calibrators. For overhead
considerations, no nodding sequences were recorded.

It immediately appeared that Hen~2-113 was fully resolved with the 250~mas
resolving power of a single-dish 8 m telescope. A few attempts to detect fringes
on some emerging structures were conducted without success.

\begin{table}[h]
\vspace{0.1cm}
 \caption{Journal of observations with MIDI/UT2-UT3: acquisition images}
 \begin{tabular}{llcrr}
 \hline\hline
 Star & Name & Time &Frames & t$_{exp}$ \\ 
 \hline

HD152786&PSF01&04:12:39 & 10000 &100s\\
HD152786&PSF02&04:13:41& 5000&50s\\
HD152786&PSF03&04:14:36& 15000&150s\\
HD152786&PSF04&04:16:28& 15000&150s\\
HD152786&PSF05&04:17:50 &15000&150s\\
Hen2-113&HEN01&04:50:46&
2000 & 20s\\
Hen2-113&HEN02&04:56:44&
2000 & 20s\\
Hen2-113&HEN03&04:59:50&2000 & 20s\\
Hen2-113&HEN04&05:02:12&2000 & 20s\\
HD152786&PSF06&05:55:34& 5000&50s\\
HD152786&PSF07&05:56:22 & 15000&150s\\
HD152786&PSF08&05:57:11 &
2000 & 20s\\
HD152786&PSF09&05:57:58&2000 & 20s\\
HD152786&PSF10&05:58:53&2000 & 20s\\
HD152786&PSF11&09:28:54& 5000&50s\\
HD152786&PSF12&09:29:45 & 15000&150s\\

 \hline
  \end{tabular}
\label{tab:journal}
 \end{table}

Custom software written in the IDL language was developed in order to
reduce MIDI observations, including images and spectra extraction.

The first step of the reduction is to read in the acquisition
datasets, average the frames on the target and the frames on the
sky, and subtract the averaged sky-frame from the averaged target-frame.
Despite the high number of optical elements in the
VLTI/MIDI system (33 in total), the quality of the 8.7~$\mu$m
images is comparable to the best MIR images published to date, i.e. Eta
Carinae (Chesneau et al. 2005).
The angle subtended by a pixel on the sky is approximatively
98~mas and the North direction rotates during the observation.

Since MIDI is a long baseline interferometer, most of the targets
are usually unresolved by a single 8~m telescope providing a wealth of instrumental PSF acquisitions.

The airmass of the targets ranges between 1.2 and 1.8, the optical seeing was
oscillating between 0$\farcs$65 and 0$\farcs$85.

The spatial resolution has been slightly increased by performing a
deconvolution using 30 iterations of the Lucy-Richardson algorithm
and the result is shown in Fig.~\ref{fig:compimage}. The spatial
resolution reached after the treatment is about 150~mas.
The levels where the different deconvolved images begin
to disagree are between 1\% of the
maximum flux of the image, depending on the quality of the
recorded images. Due to different optical properties of the beam between UT2
and UT3, only UT3 best images were used.

\subsection{ISAAC imaging}
\begin{figure}
\centering
\resizebox{\hsize}{!}{\includegraphics*[55, 355][475, 704]{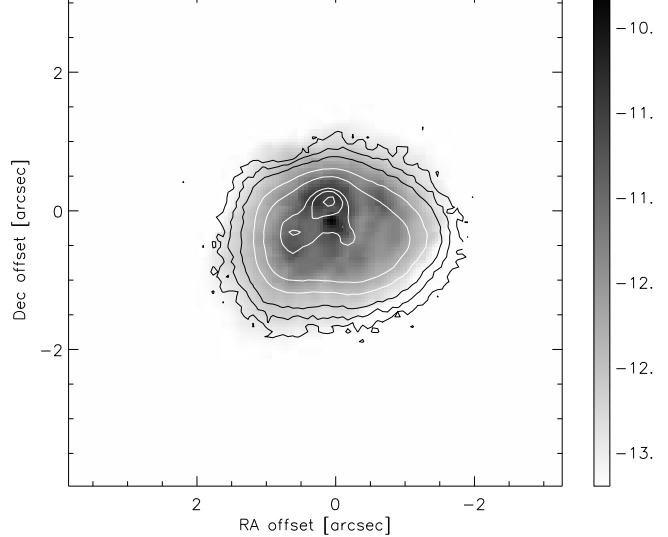}}
\caption{HST F656 image and ISAAC 3.3$\mu$m PAH contour after continuum subtraction
($\NBp-\NBpc$). North is up and east to the left.
}
\label{Fig-PAH}
\end{figure}

Near-infrared images of Hen~2-113 were taken with the Infrared
Spectrometer And Array Camera (ISAAC) at the VLT. The data
were acquired over two nights in 2004, August 05 and 06 with UT1.
The observations are summarized 
in Table~\ref{table-seeing}. Calibration 
data were obtained with short exposures of Hen~2-113 and the photometric standard
star HIP76074. The images shown in this paper were obtained on the 5th of August,
and total fluxes were scaled to be consistent with those of
the 6th of August data.
The background was subtracted by chopping and nodding to a
position 20$\arcsec$ to the north/west, while keeping the source on the
array. Jittering was used to improve flat-fielding and to minimise
the influence of the pixel-to-pixel variation. 
The pixel scale is 0.071~$\arcsec$.
The resolution in the infrared is in general better than in optical by about 25\% (Matsuura et al. 2004). 
We used {\it Eclipse 4.1.2} for the data-reduction.
A detector's non-linearity correction is adopted
as suggested by ISAAC instrument team.


\begin{table}
\begin{caption}
{Filters and image quality and fluxes.
The central wavelengths ($\lambda_0$) and the widths of the filters
are  listed for ISAAC.  
$T_{\rm{exp}}$ is the exposure time. 
Infrared seeings are measured FWHM of the nearby star
(2MASS 14595478-5417453, USNO 0300-22661298).
}\label{table-seeing}
\end{caption}
\begin{tabular}{l r@{.}l r@{.}l rc ll  r@{.}l  lll  }\hline\hline 
Band  
& \multicolumn{2}{c}{$\lambda_0$}    
& \multicolumn{2}{c}{$\Delta \lambda$}  
& $T_{\rm{exp}}$ 
&                   {Infrared seeing}  \\

& \multicolumn{2}{c}{[$\mu$m]} & \multicolumn{2}{c}{[$\mu$m]} 
& [min] &  [arcsec]                                \\

\NBpc  &  3&21  & 0&05 & 12 & 0.40   \\
\NBp   &  3&28  & 0&05 & 20 & 0.40    \\
\hline
\end{tabular}
\end{table}

Three nearby stars were used to calibrate the astrometry
by comparing to their 2MASS positions 
(2MASS 14595265$-$5417418, 14595364$-$5417517, 14595436$-$5418162).
The positions are consistent within 0.6~arcsec.
The zero-point of the figures is
RA= 14\fh59\fm53\fs4,  Dec=$-$54\fdg18\farcm7\fs0 ($J2000$)

For comparison with an optical image, we used  H$\alpha$ data obtained
by Sahai et al. (2000) (Fig. \ref{Fig-PAH}). Distortion was corrected by the
procedure of Anderson \& King (2003), and astrometry was performed using
coordinates of two nearby 
stars (14595364$-$5417517, 14595436$-$5418162).
Even after the shift of HST and ISAAC images,
there is an error of two ISAAC pixels found in the north/south direction.
For image presentation, we used pipeline reduced data,
and shifted the image according to the location of the central star.
Image distortion is negligible within the central region of Hen~2-113.

   \begin{table}
      \caption[]{Journal of observations with TIMMI/3.6m.}
         \label{tab_log}
         \begin{tabular}[]{l l l l l l }
            \hline
            \hline
            \noalign{\smallskip}
            Object &  Filter & $\lambda_0$ & Filter width &$T_{exp}$
            (s) \\
 & &($\mu m$)  &($\mu m$) & \\
           \noalign{\smallskip}
            \hline
            \noalign{\smallskip}
  Hen\,2-113           & N2  & 9.78 & 9.14-10.43 &124.32 \\
  Hen\,2-113           & N3  & 12.55 & 11.80-13.30 &62.16\\
  Hen\,2-113          & SiC & 11.65 & 10.30-13.00 &62.16\\
  HR 4763             & N2 &9.78  &9.14-10.43 &101.92\\
  HR 4174             & N3 & 12.55&11.80-13.30 &62.16\\
  HR 4174            & SiC & 11.65& 10.30-13.00&62.16\\

            \hline
         \end{tabular}
   \end{table}


\subsection{TIMMI imaging}

Observations were performed in 1995, February 05 and 07  at the ESO
3.6-m telescope in La Silla (Chile), using the  mid-infrared TIMMI 
camera (Lagage et al. 1993) with a detector consisting of  a $64\,\times\,64$
pixel Ga:Si array.
The resulting spatial scale of the system is 0.336\,\arcsec\,/pix.
The log of the observations is given in Table \ref{tab_log}. These observations
were made in a standard MIR observing mode, by
chopping the secondary mirror  and nodding the primary
to subtract the background emission from the sky and telescope. The
chopper throw was 18.3\,\arcsec \, toward the south and the nod beam position
used was 18.3\,\arcsec \, north of the first position. To avoid the
saturation of the detector by the ambient photon background and 
to have a good image quality, each individual nod cycle was split into
many short exposures of  $\sim$\,10\,ms. This procedure was repeated
for as many cycles as needed to obtain the required total integration
time. Nearly diffraction-limited images ($\sim\,0.7\,\arcsec \,$ FWHM for point
sources) resulted from these short exposures. Filter wavelengths
were selected in order to obtain information on the dust continuum as well as on
spectral features like PAHs,
while maximizing the detection sensitivity. The observations were
carried out  in four narrow-band filters, centered respectively at
 9.78\,(N2), 11.65 (SiC)  and 12.55\,$\mu$m\,(N3)
whose characteristics are given in Table \ref{tab_log}. 

The data reduction was performed using IDL self-developed routines (Lagadec et
al. 2005).  
Individually-chopped  frames were  spatially oversampled
by a factor of 4 and shifted to the nearest 0.25 pixel by using a cross-correlation 
algorithm to correct for turbulent
motions and flexure drifts.
Images were then co-added to produce a single 
flat-field-corrected image, comprising the average of the chop and nod 
differences, for each filter. Reference stars were 
observed, analyzed in the same way and were used to derive the instrumental PSF at each filter. Images were deconvolved by 
using the Richardson-Lucy algorithm. 



\subsection{HST imaging}
In order to compare our infrared images with optical images, a total of six H$\alpha$ exposures of Hen~2-113 
(F656N ($\lambda$= 656.2 nm, $\lambda$ = 2.2 nm), taken as 
part of an HST SNAPshot imaging program (GO program 8345, Sahai et al.), were retrieved from HST archives and 
processed via the standard calibration pipeline. The nebula lies entirely within the field of the 
Planetary Camera (800 $\times$ 800 pixels; pixscale = 0.0456$\arcsec$ pixel$^{-1}$) of WFPC2. 
The image has then been rotated in order for comparison  to the other
instruments.

These images (Fig. \ref{fig:compimage}) have been extensively described in
SNW00 and will be used in the following section for comparison with our observations.  

\section{Optical versus infrared morphology.}
\begin{figure}
  \begin{center}
\includegraphics[width=9.5cm]{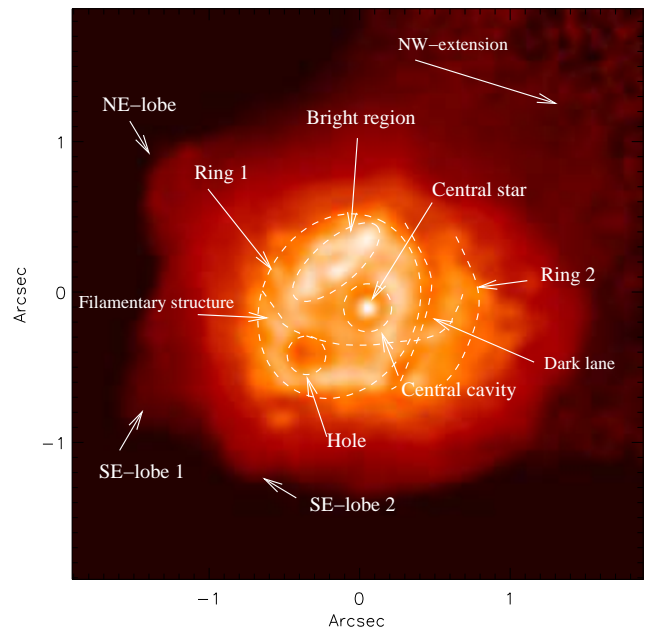}
  \end{center}
 \caption[]{Schematic view of Hen 2-113 overlaid on the Naco L' image. North
   is up and east to the left.
\label{fig:filament}}
\end{figure}


The optical nebula exhibits a bipolar shape (PA=136$^\circ$). The diffuse lobes have a radial extent of about 2"
and are well defined with sharp edges.
Superimposed on this diffuse environment, a brighter region is composed of two elliptical
rings. The second ring (ring 2) is fainter and incomplete, but its structure
is well-defined and it appears coaxial to the first one (ring 1).
The central star of Hen 2-113 is conspicuously offset from the geometrical
centers of the rings and is also not located symmetrically between these
centers. It is also offset from the midpoint of the line joining the tips of the
main NW-SE lobes of the nebula. (See Fig 1a, 1c and 3 in SNW00)

The NACO and MIDI (Fig. \ref{fig:compimage},  \ref{fig:L_image},
\ref{fig:filament} and \ref{color}) high resolution images show a nebula whose extension is limited to the bright core of HST images.
The limits of this core are as reported by De Marco et al. (1997) about 1.4\arcsec x1.1\arcsec.

The star, visible in the L'
and M' NACO images disappears in the N band (MIDI $8.7 \mu$m). A bright ring-like
knotty structure which coincides with ring 1 observed by
SNW00 is also observed in the L', M NACO images, and in the
MIDI 8.7$\mu$m deconvolved image. This ring is roughly elliptical and
  the central star is offset from the geometrical center of the ring of
  $0.15\arcsec$ along P.A$\sim65^\circ$. Its major axis has
  an extension of $\sim 1.2\arcsec$ and a minor-to-major axis ratio of
  $\sim0.6$ and the line passing through the center of the ring and
  perpendicular to its major axis is orientated at a P.A. of $\sim65^\circ$. Ring 2 is less clearly
visible and is replaced by an elongated region of diffuse
emission from L' images to N band images. These rings are thus smaller
  in our mid-infrared observation (Fig \ref{fig:compimage}) than the one
  observed with HST. The three color composite image (Fig. \ref{color}) indeed shows that the
  rings observed at mid-infrared wavelenghts lie inside those observed in $H_{\alpha}$.

The dark lane ($0.2\arcsec$ width in L' and HST images) in the L' images which splits the
emission between the western border of the bright ring and this
elongated diffuse emission is coincident with the HST
one (Fig. \ref{color}). The lane can be distinguished in the M' and 8.7$\mu$m images
but the reduction of the spatial resolution at these wavelengths together with the decrease of
optical thickness prevent from a clear identification of this structure. Such a
dark lane suggests the presence of a ring of cold dust.

A bright region is observed on the North-East of ring 1 in our NACO L'
  and M' image as in MIDI $8.7\mu$m image. This blob may represent
  emission from dust in excess in this part of ring 1. It is 2 or 3
  times brighter than the other parts of ring 1 in $H_\alpha$, L', M' and MIDI
  images and its dimension is roughly $0.7\arcsec \times 0.2\arcsec$.

A small circular region of low emissivity (hole), probably due to a lack of material, is
  detected at $\sim 0.5 \arcsec$ to  the South-East of the central star along a
  P.A.$\sim 128^\circ$ in our two NACO images, and can be distinguish with the
  HST and MIDI observations. Its diameter is roughly $0.3\arcsec$.
It is interesting to notice that this direction nearly coincides with the
orientation of the bipolar nebula surrounding the two rings as observed by
SNW00 (P.A. $=136^\circ$). 
Furthermore, our deconvolved NACO L' image seems to show the presence of
material on the North-West, alignated with  the direction formed by the
central star and the hole. Another possible interpretation would be the
presence of a fairly dense clump of dust as observed in the PN Roberts 22
(Sahai et al. 1999). Nevertheless in this case we wouldn't expect to see the
bright blob in $H_{\alpha}$ as seen in HST image.

More complex structures are seen in the high dynamics L' NACO deconvolved image
(Fig. \ref{fig:L_image}). In particular ring 1, the lane  and ring 2 seem to
be broken by a thin emissive filamentary structure  which can be traced
to the eastern rim of ring 1, passing $\sim 0.2\arcsec$ south from the
star. This $0.1\arcsec$ width filamentary structure is roughly
  perpendicular to the dark lane.

This image also shows that the bright core (the two rings and the lane)
  is surrounded by a faint low emission region of $2.3\arcsec \times
  3\arcsec$.  The brightest region of ring 1 is one hundred times brighter than
  this "halo".
  Three lobes are seen extending from this region. The brightest (NE-lobe) is
  orientated along a P.A. $\sim 55^\circ$, coincident with the NE-lobe described by
  SNW00, supporting the reality of the fine structures observed in
  our deconvolved NACO image. Another lobe (SE-lobe 1) is observed in the SE direction,
  oriented along P.A. $\sim 115^\circ$. The counterpart of this lobe could be the
  material seen on the north/west. The smallest lobe (SE-lobe 2) is also orientated roughly SE, with a
  P.A. $\sim142^\circ$. 

A diminution of  dust emission is detected close to the CS (Figs
\ref{fig:radcavity}, \ref{fig:L_image}, \ref{fig:filament}), which is placed in a cavity of radius
$\sim 0.3\arcsec$. Such a cavity is not seen in the H$\alpha$
images obtained with the HST, suggesting that ionized gas fills the cavity or is seen
in front of it.

\section{Interpretation}

\subsection{The structure of Hen2-113}
\label{sec:struc}
  \begin{figure}
  \centering
 \includegraphics[width=8.5cm]{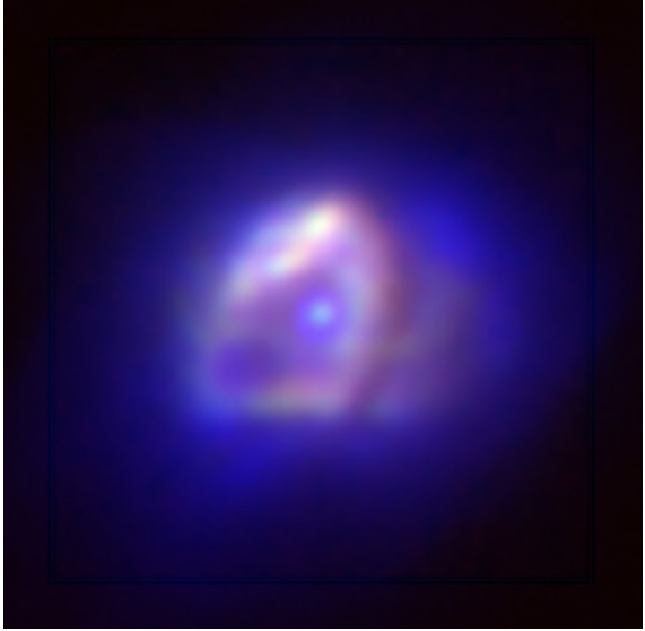}
   \caption{Three color composite image of Hen2-113 made by combining  HST $H_{\alpha}$ 
 (blue),  NACO L' (green) and M'(red) images. North is up and east to the left.}
   \label{color}
  \end{figure}

HST observations (SNW00), show that Hen~2-113 has an overall bipolar shape and two 
ringlike structures around the CS. SNW00  noticed also that the CS
was offset from the geometrical center of the
rings. This is confirmed by our NACO and MIDI observations and
motivated us to make a simple geometrical model to explain it.

\begin{figure}
  \centering
  \includegraphics[width=8.5cm]{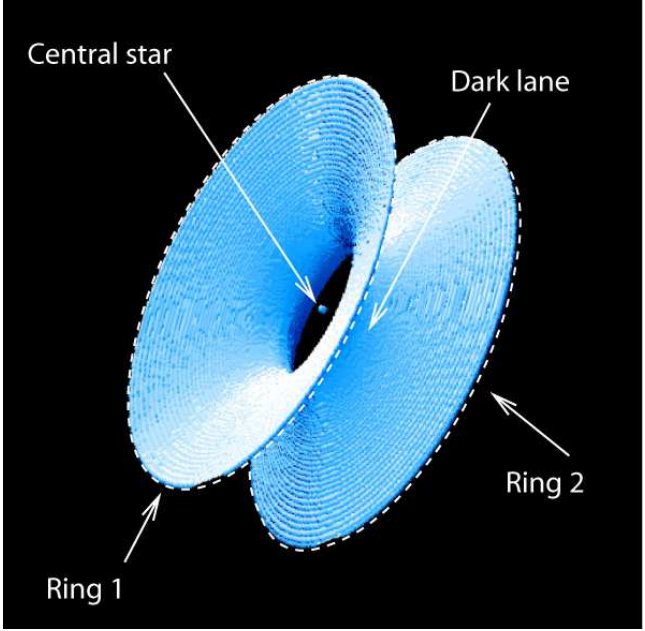}
   \caption{Diabolo model with the adopted inclination parameters}
   \label{diabedge}
  \end{figure}

To disentangle projection effects and to understand the real three dimensional structure 
of an astronomical object is always a
difficult task. First, we notice that ring 1 (eastern ring) is brighter than ring
2 (western ring), suggesting that ring 1 is closer than ring 2. A three color composite 
image (Fig. \ref{color}) seems to confirm this hypothesis. Indeed, the superposition of our L' and M'
images with the HST $H_{\alpha}$ image indicates that the two rings
are the projection of a diabolo-like torus (Fig \ref{diabedge}), with the eastern
ring (top of the torus) pointing toward us. In such a geometry, the brightness
difference of the two rings is easily explained by a screening effect from the
equatorially enhanced dust density. 
The ``diabolo-like" surface is represented by a revolution hyperboloid defined by the equations:

$x=a (1+u^2)^n \cos v$

$y=a (1+u^2)^n \cos v$

$z=c u$

with: -$\frac{\pi}{2}\leq u \leq \frac{\pi}{2}$ and\, -$\pi \leq v \leq \pi$.

The parameter $a$ fixes the radius of the diabolo, $c$ is the height and $n$ the opening angle. 
The code, written in IDL, produces an hyperboloid, which can be scaled
according to its distance and inclination with respect to the plane of the sky, allowing
a direct comparison with the observed images. The fit of the parameters was performed visually,
searching for the best superposition between the simulated and the observed
images, the HST optical image being the  reference, as ring 2 is best
seen on  HST $H_{\alpha}$. The aim of  this model is to reproduce the main morphological features
observed for the rings: the major-to-minor axis ratio of ring 1 is $\sim$ 2,
two thirds of ring 2 is observed, the central star is offset with respect
to the center of the rings and the projection of the common axis of these
rings on the sky-plane is oriented with a P.A. of $\sim65^\circ$.

Fig \ref{diabedge} shows a representation of our model. A satisfying superposition
was obtained with a diabolo having an inclination $i=37^\circ$  and a P.A
  $\sim 6
5^\circ$, $n=0.9$ and $a/c=3/2$. This geometrical model can explain the
offset of the CS from the center of ring 1, but the offset from the center of the 
nebula (SNW00) is not explained by this geometry. It can be explained
by proper motion of the star at $\sim1$ km s$^{-1}$ with respect to the nebula. 

The diabolo-shaped structure is
tilted with respect to the bipolar nebula observed by SNW00, oriented along
P.A.$=136^{\circ}$. Notice that the orientation of the diabolo
(P.A.$\sim65^\circ$) is coincident with none of the other features observed inside
the nebula. 

SNW00 already noticed that the shaping of this nebula could not be
explained with the generalized interacting stellar wind model (GISW) (Kwok et al. 1978, Balick
1987) as this model cannot explain multipolar morphologies. 
They proposed that the observed morphology could be explained if fast,
collimated outflows change their direction with time (Sahai \& Trauger 98, hereafter ST98).
Our observations  strengthen the fact that the observed morphology cannot be
explained by the GISW model as the morphology of the nebula is
multipolar. The observed tilt also seems to be difficult to explain
by the ST98 model. The hole observed in the south/east of the nebula
could be a clue to our understanding of this structure. This hole is probably due to a 
lack of dust in this part of the nebula and it is interesting to notice that the direction formed by the
CS and the hole (P.A.$\sim 128^{\circ}$) is nearly coincident with
the orientation of the main bipolar nebula (P.A.$\sim 136^{\circ}$). This is
seen clearly in our three color image (Fig. \ref{color}).

Since the nebula around Hen~2-113 is relatively young, it is only partially ionized, as
the ionization front is still propagating inside the shell. The total measured H$\beta$ flux (Acker et al. 1992) 
corrected by an extinction E(B-V) = 1.0 (De Marco, Barlow \& Storey 1997) is 
$F(H\beta) = 4.08\times 10^{-11}\,erg.cm^{-2}.s^{-1}$. In this case, the ionized mass is given
by
\begin{equation}
M_{ion} = 4 \pi D^3\frac{F(H\beta)m_p}{j(H\beta)<n_e>} = \frac{31D_{kpc}^2}{<n_e>}\, M_{\odot}
\end{equation}
where D is the distance, $j(H\beta)$ is the line emissivity coefficient, $m_p$ is the proton mass
and $<n_e> = 10^5 cm^{-3}$ is the mean electron density in the nebula (de Freitas Pacheco et al. 1992).
For distances in the range 1-3 kpc, the ionized mass is in the range $(0.3-2.8)\times 10^{-3}\, M_{\odot}$.
For PNe, the mass of the ionized shell is anti-correlated with the mean electron density (Pottasch 1984).
This can be understood by the fact that, as the ionization front propagates increasing the ionized mass,
the shell expands and the density decreases. For a young nebula with a density of about $10^5\, cm^{-3}$,
the expected ionized mass is about $(2-4)\times 10^{-3}\, M_{\odot}$ (Pottasch 1984), consistent with
our findings and supporting the idea that Hen~2-113 is a young nebula. Note that SNW00 estimated that
the expansion age of Hen~2-113 is $<$ 550 yr.

\subsection{Infrared excess from circumstellar material close to the CS}

As mentioned in Section 2.2, the photometry of the CS gives $m'_{L'}$=8.4$\pm$0.1 mag
 and $m'_{M'}$=8.3$\pm$0.2 mag for
 filters L' and M', respectively. If we compare these values with predictions from atmosphere models, it 
appears that a star with the same effective temperature as Hen~2-113 should
 not  be so bright at those
wavelengths.


Indeed, using Kurucz models (Kurucz 1979) with $T_{eff}=29000$ K (De Marco \& Crowther 1998), we have
estimated L' $\sim$ 14.5 and M' $\sim$ 15.5, by adopting J = 9.78 (Webster and Glass 1974, 2MASS=9.51$\pm$0.026) and assumining
no extra light contribution to the CS emission at this wavelength. This assumptiandon seems to be justified by
the analysis of continuum emission from the UV to the far-IR (see Fig.1 by De Marco \& Crowther 1998).
Therefore, the observed flux from the central object in the L' and M' bands is about $\sim$300 and $\sim$800 
times  higher respectively than expected from a model including only the central star. 
It is worth mentioning that a comparison between the PSF and the light profile of the bright central 
object indicates that the latter is resolved by our data, having an angular diameter of about
155 mas in the L' band, corresponding to dimensions of 140-430 A.U., if the distance of Hen~2-113 is in
the range 1-3 kpc. Notice that a similar infrared excess was also observed by Matsuura et al. (2004) associated to
the emission of the CS of NGC~6302.

We tested several hypothese to explain this small and intense emission, in particular free-free and dust emission.

In the hypothesis that the observed infrared excess results from bremsstrahlung, the required ionized gas mass within
the central region is $M\sim
7 \times 10^{-5}-
4 \times 10^{-4}M_{\odot}$ (adopting $T_e$=10000 K and $n_e=10^5$cm$^{-3}$, de Freitas Pacheco et al. 1993) comparable to the present mass of ionized gas
estimated for the whole nebula in Sect.~\ref{sec:struc}. Thus this hypothesis
seems implausible.

We then considered the possibility that the infrared excess is produced by a
``cocoon'' of hot amorphous carbon dust.

The wind of the CS of Hen~2-113 is C-rich and one could expect the formation
of carbon dust grains
as the envelope cools in the expansion and is decelerated by the material previously ejected. If
re-radiation by dust is the origin of the infrared excess observed in the L' and M' bands, the
mass of grains $M_d$ inside a radius of 70-210 A.U., within the optically thin approximation, is
\begin{equation}
M_d = \frac{4D^2f_{\lambda}a\delta}{3Q_{\lambda}B_{\lambda}(T_d)}
\label{dustmass}
\end{equation}
where $f_{\lambda}$ is the observed flux in L' (or M') band, $a$ is the mean grain radius, 
$\delta$ is its density and $Q_{\lambda}$ is the absorption(emission) efficiency factor.  

The grain temperature, $T_d$, can be estimated from the flux ratio at two different wavelengths
(3.8 $\mu$m and 4.78 $\mu$m in the present case), e.g.,
\begin{equation}
\frac{f_{\lambda_1}}{f_{\lambda_2}} = \frac{Q_{\lambda_1}B_{\lambda_1}(T_d)}{Q_{\lambda_2}B_{\lambda_2}(T_d)}
\end{equation}

Assuming dust grains of amorphous carbon ($\delta = 2.0$ ~g ~cm$^{-3}$) with $a$ = 0.2 $\mu$m, a temperature 
of $\sim$ 1000 K is derived from the equation above. This high temperature would be expected for 
carbon grains close to the central star, not very far from the inner regions hotter than 1500 K, when 
sublimation of amorphous carbon occurs. Once the temperature is known, the dust mass can be estimated from
Eq.\ref{dustmass} and, for the range of distances, D, considered, one obtains $M_d \sim (7-28)\times 10^{-10}\,
$M$_{\odot}$ and gas masses about two orders of magnitude higher\footnote{This a lower limit for the mass. The physical conditions encountered so close to the CS would probably
lead to a dust-to-gas mass ratio lower than 1\%.}. The required dust masses imply optical depths of
about $10^{-3}$ around 4$\mu$m, justifying our optically thin approximation. 
These values, which represent a few years of mass-loss, are not in conflict with the nebular
ionized gas mass estimated in in Sect.~\ref{sec:struc} and the wind crossing
time of the small circumstellar region considered is also a few years.

We note that the temperature and dust mass estimated here are remarkably
consistent with the mass and temperature independantly estimated by SNW00
for their hot dust component by SED fitting and by De Marco \& Crowther (1998)
by fit to the IR excess.

\subsection{PAHs}
  \begin{table}
      \caption[]{Results of the fits to our image with a 2-dimensional
        Gaussian ellipse. $a/b$ is the ratio between the long and short axis
        of the ellipse, and $\theta$ is its orientation .}
         \label{table_gauss}
         \begin{tabular}[]{l c c c}
            \hline
            \hline
            \noalign{\smallskip}
            Filter & $a/b$ & P.A.     ($^\circ$) & PAH content (\%)\\
           \noalign{\smallskip}
            \hline
            \noalign{\smallskip}
  $8.7 \mu$m       & 1.09    & 68 & 37\\
  N2               & 1.05    &  -   & 4\\          
  SiC              & 1.16    & 91 & 20 \\
  N3               & 1.05    &  -   & 11 \\
  NB$_-$3.21       & 1.06    & 91   & -  \\
  NB$_-$3.28       & 1.16    & 91 & 54  \\
  Isaac subtract.  & 1.34    & 92 & 100\\
         \noalign{\smallskip}
            \hline
         \end{tabular}

\small{ The
  orientation having no meaning for spherical objects, we did not mention
  angles for N2 and N3 filters.}

   \end{table}
Hen~2-113 belongs to a group of bipolar post-AGB stars at the center of young
ionised PNe showing both PAH bands and
crystalline silicates.
A scenario proposed by Waters et
al. (1998) and Molster et al. (1999) explains the observed dual chemistry
observed around these stars. In this scenario, a disk, formed while the star was still
oxygen-rich, remains during the high mass-loss phase when the
star is carbon-rich, forcing the matter outflow to a bipolar geometry
perpendicular to the direction of the disc.

Thus, we expect to observe different spatial distributions for PAHs and O-rich grains. Unfortunately, O-rich dust spectral features are
  observed at wavelengths $>$ 15 $\mu$m where  atmospheric absorption makes
  ground-based observations very difficult. Observations at longer wavelengths
  (space-based, or with new MIR detectors) where the
  O-rich dust features lie would certainly bring new insights on the
  spatial distribution of O and C-rich dust.

Mid-infrared spectra of Hen~2-113 display strong features that can be attributed to PAHs.
Different filters were used to observe a varying and significant amount of emission from these PAHs.
The underlying continuum was estimated based on the work of Cohen et al. (2002), who represented the data 
by the superposition of a blackbody having a temperature of 395 K and that
of a grey body ($\lambda^{-1.2}B_{\lambda}(T)$). The results by Cohen et al. are comparable to 
those by SNW00, who fitted the continuum by three blackbody curves,
cold  ($T_c$ = 100 K), warm  ($T_w$ = 378 K) and hot
($T_h$ = 900 K). We retrieved the ISO spectra of Hen2-113 from
the archives and substracted it from the  continuum modeled by two blackbodies. We then assume that the features resulting from this
substraction was due to PAH. We then compare this with
the filter transmissions to derive the PAH contribution in each filter.

The estimated PAH contributions in the different filters are 54\%,37\%, 4\%,
20\% and 11\% for ISAAC $3.28\mu$m, N8.7$\mu$m, N2, SiC and N3, respectively. These estimates
are based on the assumption that the ISO fluxes recorded with large apertures
are equivalent to the  fluxes recorded with
the various instruments used in this work.
To study the
morphology  of Hen~2-113 in different filters quantitatively  we fitted our
images with a 2-dimensional Gaussian function. The best-fit parameters
are summarized in Table.~\ref{table_gauss}. The observations in the PAH bands
(ISAAC $3.28\mu$m, MIDI
$8.7\mu$m and TIMMI SiC filter) have a stronger major-to-minor axis ratio and tend to be 
oriented roughly east/west (except the MIDI image which has a P.A.$\sim68^\circ$). 

To check if the PAHs and dust grains responsible for the
  continuum emission have  different spatial distributions, we compared
  observations made with filters in the continuum and filters where the PAH
  contribution is expected to be significant. We thus compared  ISAAC $3.21\mu$m (continuum)
  and $3.28\mu$m (PAH contribution $\sim 54\%$) images as well as TIMMI N2
  (continuum) and MIDI $8.7\mu$m (PAH contribution $\sim 37\%$)
  observations. Note that MIDI and TIMMI have different resolutions, so the
  MIDI observations were artificially degraded to the TIMMI resolution
  ($\sim0.7\mu$m) by simple convolution with a 2D gaussian to make these
  observations comparable.

  Both ISAAC images show a single bright blob at the centre and a halo
  surrounding the center. Both  blobs have an elongation toward the south/east
  and south/west direction. The halo is almost symmetric with a slight
  elongation, a bright blob at the center and a nearly spherical halo. The
  TIMMI N2 (continuum) and MIDI 8.7$\mu$m images also have a
  similar morphology. However, we can note that on the N2 image the object is
  broader than on the
  8.7$\mu$ image. Whether this is due to the fact that the observations were
  made with different instruments and exposure times or wether the dust grains
  responsible for the continuum emission have a broader spatial distribution
  than PAH dust grains is hard to check. 

Thus, we focused our work on the
  comparison of both ISAAC images by comparing radial cuts obtained with
  different filters (PAHs and continuum). First, using azimuthal averaging of
  these radial cuts, we note that in the PAH image the object is broader than
  on the continuum image. We then tried to check if there was some preferential
  direction for the broadening of the PAH image. We find that the strongest
  asymmetry (i.e., the intensity ratio for which a given radius is maximum) is
  observed for a P.A.$\sim139^\circ$ , the intensity ratio being
  $\sim1.55$. The strongest mean asymmetry (i.e. mean value of the intensity
  ratio) is observed for P.A.$\sim248^\circ$ (the intensity ratio is $\sim1.24$).

 Finally, our ISAAC observations show that PAHs  seem to have a broader spatial
  distribution than the grains
  responsible for the continuum emission. As shown in Sect.~\ref{sectpahmor}
  the largest difference between the ISAAC PAH and continuum images are observed
  for P.A.$\sim139^\circ$ and $\sim248^\circ$. It is interesting to notice
  that these orientations correspond to the orientations of the  main bipolar nebula and the side of
  the diabolo-shaped structure  pointing away from us. These structures being
  the youngest structures of the nebula, this tends to confirm the fact that
  the PAHs have been formed more recently than the dust grains
  responsible for the continuum emission.
\section{Conclusion}
In this paper, we report high-spatial resolution infrared observations of the young PN
Hen~2-113 obtained with different instruments at ESO, Chile (NACO, MIDI, ISAAC
and TIMMI). These observations provide new insights on the complex morphology
and structure of the nebula around it. 

These observations, thanks to the high resolution obtained with adaptive
optics on VLT (NACO), led to the discovery of features previously
unseen in this nebula and which seem  difficult to explain with current
models for the shaping of PNe. We discovered a void of $\sim0.3\arcsec$
in diameter around the central source and a hole in the
South-East of the nebula. The fact that the direction formed by the CS and
this hole is nearly coincident with the orientation of the main axis of the nebula could
be a clue for the understanding of the structure of the nebula. A simple
geometrical model indicates that the CS is surrounded by a diabolo-shaped
dusty structure which is tilted with respect to the rest of the nebula.

Our infrared data indicate an infrared excess with respect to the expected
stellar emission based on stellar models and shorter wavelength data. We tested different
hypothesis to explain this infrared excess. Dust
emission very close to the CS seems to be responsible for this excess. Indeed,
simple calculations indicate that emission from
hot dust (T$\sim 900-1000$K) with mass $\sim 10^{-9}$M$_{\odot}$ can account
for the infrared excess. The fact, that in their modeling of the SED of
Hen~2-113, SNW00 need a hot dust component with the same mass and same
temperature, strengthens the hypothesis that the infrared excess is due to
emission from hot dust. We also note that such an infrared excess has
been observed in at least one similar object (NGC 6302 (Matsuura et
al. 2005)), indicating that this could be a feature common to this class of object. We would
thus need  high-angular resolution observations of other PNe to confirm this
hypothesis.

These infrared observations, obtained with different instruments and wavelengths,
allowed us to study the spatial distribution of PAHs in the nebula of
Hen~2-113, known for displaying strong features attributed to PAHs. These
observations indicate that the images made with filters containing PAH
features seem to be broader than the one observed with continuum filters. The
ISAAC observations indicate that the  PAH/continuum difference is stronger
along the direction of the main nebula and the diabolo-shaped torus. However,
it is difficult with these observations to verify the scenario proposed by Waters
et al. (1998) and Molster et al. (1999) to explain the dual chemistry in this
class of nebulae. Observations at longer wavelengths ($>15\mu$m), where
spectral features of O-rich dust are observed, would thus be necessary to compare
the spatial distribution of PAHs and O-rich dust.

\begin{acknowledgements}
The authors warmly thank those who made possible these service mode observations, in particular
the ESO  night astronomers for their careful work.
OD would like to acknowledge Janet Jeppson Asimov for financial support.
\end{acknowledgements}

\end{document}